\documentclass[12pt]{article}
\usepackage{geometry}                
\usepackage{graphicx}
\usepackage{amssymb}
\usepackage{cite}
\usepackage{amsmath}
\usepackage{bm}

\newcommand{\beq}{\begin{equation}}
\newcommand{\eeq}{\end{equation}}
\def\beqn{\begin{eqnarray}}
\def\eeqn{\end{eqnarray}}
\def\lsim{\mathrel{\rlap{\lower3pt\hbox{\hskip0pt$\sim$}}
    \raise1pt\hbox{$<$}}}
\def\gsim{\mathrel{\rlap{\lower4pt\hbox{\hskip1pt$\sim$}}
    \raise1pt\hbox{$>$}}}

\begin{document}

\title{~\\
\bf 
Highly Excited Mesons, Linear Regge Trajectories
and the Pattern of\\ the Chiral Symmetry Realization
 }
\author{M.\ Shifman and A.\ Vainshtein\\[2mm]
{\small \sl William I. Fine Theoretical Physics Institute},\\[-1mm]
{\small \sl University of Minnesota,
Minneapolis, MN 55455}
}
\date{}                                           

\maketitle
\thispagestyle{empty}

\vspace{-9.5cm}

\begin{flushright}
FTPI-MINN-07/26\\ 
UMN-TH-2616/07\\
\end{flushright}

\vspace{6.5cm}

\begin{abstract}
 
The chiral symmetry of QCD shows up in the linear Weyl--Wigner mode 
at short Euclidean distances or at high temperatures. On the other hand, low-lying 
hadronic states exhibit  the  nonlinear Nambu--Goldstone mode.
An interesting question was raised as to whether 
the linear realization of the chiral symmetry is asymptotically restored for 
highly excited states.  We address it
in a number of ways. On the phenomenological side we argue
that to the extent the meson Regge trajectories are
observed to be linear and equidistant,  the Weyl--Wigner mode is 
not realized.

This picture is supported by quasiclassical arguments
implying that the quark spin interactions in high excitations are weak,
the trajectories are linear, and there is no chiral symmetry restoration. 
Then we use the string/gauge duality. In the  
top-down Sakai--Sugimoto construction the nonlinear realization of the
chiral symmetry is built in. In the bottom-up AdS/QCD construction by 
Erlich et al.,  and Karch et al. the situation is
more ambiguous. However, in this approach linearity and equidistance of
the Regge trajectories can be naturally implemented, with the chiral symmetry
in the Nambu--Goldstone mode. 

Asymptotic chiral symmetry restoration might
be possible if a nonlinearity (convergence) of the Regge trajectories 
in an ``intermediate window" of $n,J$, beyond the explored domain, takes place.
This would signal the failure of the quasiclassical picture.

\end{abstract}

 \newpage
 


\section{Introduction}
\label{intro}

Since the inception of QCD till the end of Millennium the
prime interest of the QCD practitioners was the spectrum  and properties
of the low-lying hadronic states, such as $\rho$ mesons, pions and nucleons.
A number of methods was developed to treat such states,
starting from the soft-pion technique which predates QCD by a decade,
then QCD sum rules, lattice calculations and so on. Little attention was paid to highly excited states. The reason is obvious: the decay widths of the excited states grow
with the excitation number, so that they overlap and collectivize themselves, and
could be treated as  continuum.

In the Regge theory which dominated high energy theory
before  QCD, highly excited states played an important
role in phenomenological analyses since they determine the daughter Regge trajectories.
The Regge theory gave rise to dual resonance models
which eventually grew into string theory.
Ironically, string theory that emerged from the dual resonance models
shortly after  became ``string theory for nonhadrons,"
 and was elevated to the status of
``theory of everything" in the 1980s and early '90s.
With this promotion the previous interest to excited hadronic states
faded away.  At the same time, in QCD highly excited states were 
treated as belonging the the realm of asymptotic freedom 
which inevitably qualified them as ``dynamically  uninteresting objects."

\begin{figure}[h]
 \centerline{\includegraphics[width=0.95\textwidth]{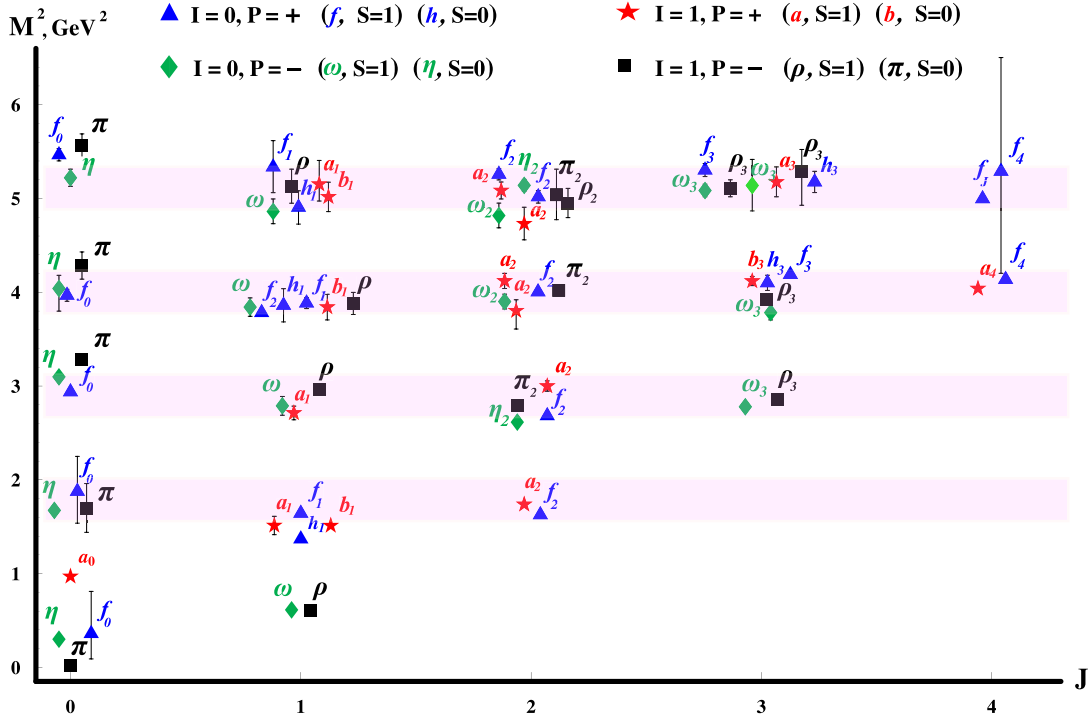}}
 \caption{\small {%
 The plot shows $M^2$ of various meson resonances  which are
 believed to be built of $\bar q q$ where $q=u$ or $d$. The resonances at 
 levels 2, 3 and  some resonances at 4 level GeV$^2$
 are taken from the Particle Data Group (PDG) compilation. Most of those at level 4
 and all resonances at level 5 GeV$^2$
 are taken from the compilation of resonances in $p\bar p$ annihilation prepared by Glozman \cite{Gloz2}, see also \cite{Bugg:2004xu}. In selecting the $\bar q q$ resonances we followed Kaidalov's work \cite{Kaidalov} 
 in discarding presumed four-quark states, gluonia
 or resonances built of $\bar s s$. }}
 \label{fone}
 \end{figure}

This attitude changed in recent years with the advent of string--gauge duality methods,
based on the 't Hooft limit \cite{hl} with the number of colors $N_{c}\to \infty$ 
while $g^2 N_{c}$ is kept fixed.
In this limit the meson decay widths tend to zero, so that
individual highly excited mesons become well-defined.\footnote{%
\,Baryons, if treated in the standard
't~Hooft procedure, defy this rule; their decay widths, generally speaking, do not
vanish in the limit $N_{c}\to\infty$, also their masses grow as $N_{c}$. 
However, the $N_{c}\to\infty$ limit  exists for the mass differences, and experiments show that rather  high excitations of nucleons and other baryons can be identified using the existing data. }

The string--gauge duality-based ideas
predict a certain pattern for excited resonances.
On the other hand, significant amount of data
regarding excited mesonic resonances was accumulated. These data
shown  in Fig.\,\ref{fone} exhibit a high degree of degeneracy.\footnote{\,We should warn the reader
that there is no consensus among experts with regards to 
some resonances at higher levels and some selection criteria
for $q\bar q$ states, see the figure caption.}
In classical strings the Regge trajectories are linear and equidistant
implying the spectral degeneracy. 
The mode of coexistence of
the chiral symmetry with the  Regge trajectories is a challenging theoretical
issue.

Generically, the chiral symmetry could be realized
linearly, i.e. in the Wigner--Weyl mode, when chiral multiplets contain 
degenerate states of the opposite parity, or non-linearly, in the 
Nambu--Goldstone mode, 
in which the action of symmetry generators adds soft pions instead. 
Of course, we know for sure from the low-lying states that the 
chiral symmetry is realized in 
the Nambu--Goldstone mode in the hadron world.
 
The question is whether the linear realization
of U$(N_{f})_{L}\times$U$(N_{f})_{R}$
could be restored for highly excited states yielding  
a part of degeneracy visible in Fig.\,\ref{fone}.  
In this case one could speak of the
asymptotic chiral symmetry restoration ($\chi$SR). On the other hand,
if in higher excitations the Nambu--Goldstone mode persists, other dynamical reasons must be responsible for the spectral degeneracy. The issue of possible restoration 
of the full U$(N_{f})_{L}\times$U$(N_{f})_{R}$ chiral symmetry 
attracted much attention lately mainly in connection with 
the inspiring works of Glozman and 
collaborators \cite{Gloz0,Gloz1,Gloz3,Gloz4,Afo,Gloz5}. 
The history of the topic of parity doubling on the
Regge trajectories is presented in the review papers \cite{Greport,safo}.

The above dichotomy --- restoration vs.\ nonrestoration --- is in the focus of the present work.
We first explain that, purely theoretically, the Nambu--Goldstone
mode of the chiral symmetry implementation for the low-lying states
could coexist with the linear realization for high excitations.
Which of the two alternatives takes place in actuality
is decided by dynamics.

In the first part of our paper we focus on high radial excitations
in $q\bar q$ mesons where $q=u$ or $d$.
We argue that the observed approximate linearity in the $\{M^{2}, J\}$ plane 
and equidistance of the $q\bar q$-meson Regge trajectories  ---
to the extent and in the domain they are observed ---
imply nonrestoration of the linear chiral symmetry. 

The genuine restoration means that the mass difference between the would-be chiral 
partners of the opposite parities is much less than, say, the gap to 
the next radial excitation. However, for linear trajectories they both scale as $1/M$. 
Under the circumstances, the very notion of the ``chiral
partners" becomes meaningless;
rather we deal with an ``extended promiscuous family." A large number of states
of positive and negative parity are connected with each other by axial transitions.
This is typical of the Nambu--Goldstone, rather than Wigner--Weyl  mode:
excited mesons are not  decoupled from the pion and the axial transitions between mesons can have arbitrary strengths.

Our conclusion is in obvious contradiction with the interpretation suggested 
in early works
\cite{Gloz0,Gloz1,Gloz3} in which
the $1/M$ fall-off of the mass splittings between the chiral partners
was considered to be a signal of the asymptotic $\chi$SR.
Theoretical basis for some of these works was provided by  
a straightforward four-dimensional extension 
\cite{Glozman:2005tq,Wagen,Greport}
of the two-dimensional 't Hooft 
model \cite{TM,TM1}.  

This two-dimensional model is fully understood. For massless quark it has U(1)$_{L}\times$U(1)$_{R}$
chiral symmetry. This symmetry is spontaneously broken by the quark condensate,
a massless Goldstone meson ensues. The fact that 
the chiral symmetry is not restored in high radial excitations (needless to say, there are no orbital excitations in two dimensions) is clearly seen from the Goldberger--Treiman relation 
$$
g_{\pi + -}=f_{\pi}^{-1}g_{A}(M_{+}^{2}\!-\!M_{-}^{2})\,,
$$
where $g_{\pi + -}$ is the pion coupling to $|\pm\rangle$ mesons of 
opposite parity, with masses
$M_{\pm}$. The excitation mass spectrum is 
known to be equidistant in $M^{2}$;  for the nearest neighbors
$M_{+}^{2}\!-\!M_{-}^{2}$ is independent of the excitation number. While the pion decouples
in the $N_{c}\to \infty$ limit, 
the coupling $g_{\pi + -}$ does not fall off with the excitation number.

Unlike the two-dimensional  't Hooft model, the four-dimensional QCD dynamics
is not fully understood. Therefore, generally speaking, 
the observed pattern of the Regge trajectories could  change  outside
 the measured domain  of $\{M^{2}, J\}$. In particular, the chiral partners could approach each other  at large $J$. 
This would certainly imply a strong nonlinearity of the Regge trajectories
at the right edge of Fig.\,\ref{fone} and beyond. 
Although data give no hint of such a behavior, logically it is not ruled out.

What does theory say on this issue?

In the framework of holographic string/gauge duality the most developed construction was suggested by Sakai and Sugimoto \cite{ss}.
Albeit this  construction does not reproduce the linearity of the Regge trajectories,
the Nambu--Goldstone mode of the chiral symmetry realization is built-in.

On the other hand,  the bottom-up AdS/QCD approach 
of Erlich et al. \cite{Erl} and Karch et al. \cite{Kar} does not lead to a unique answer.
Although in Ref. \cite{Kar} the authors obtained asymptotically linear realization
it was at a price of nonlinearity of the Regge trajectories (as a function of $n$). 
Moreover, within the same approach one can change a certain assumption 
(see, e.g. \cite{CKP, Bergman}) ensuring the linearity of the trajectories and keeping
the Nambu--Goldstone mode for high excitations. 

While the analysis based on the  string/gauge duality is still open
for interpretations, arguments based on quasiclassical considerations seem to be
much more definite. They imply that there is no $\chi$SR at high $n,\,J$. 
If  a convergence of trajectories at modestly large  values of $J$ was found
(signaling the beginning of $\chi$SR in the large-$J$ limit), this would
defy the quasiclassical picture.

A comment on baryons is in order here.
There is no clean formulation of the problem for arbitrarily
large $n$ and $J$ since, as was mentioned, the large-$N_c$ 
limit does not help for baryons:
starting from a certain value of the excitation number
they are expected to overlap and become unidentifiable.
However, given the fact that excitations of $N$ and $\Delta$
with intermediate values of $n$ and $J$ are observed,
one can pose the question of $\chi$SR with regards to  these ``intermediate"
excitations. We discuss this question in the second part of our paper.

We also discuss a relation between the chiral symmetry mode and relevant distances.
At short Euclidean separations the chiral symmetry is restored, which is obvious
from the operator product expansion (OPE). However, Euclidean considerations are
not sensitive enough to the relative positions of individual chiral partners. Looking from the Minkowski 
side, we observe the growth of characteristic  distances with energy: the meson size grows linearly with its
mass. Note that the chiral symmetry restoration  at high temperatures is similar
to that at short Euclidean distances. In both cases we average over a large number of resonances.

This paper is organized as follows. In Sect. \ref{coeng}
we discuss how the nonlinear and asymptotically linear realizations
could coexist, in principle, at low and high energies, respectively.
In Sect. \ref{qamrt} we classify the $\bar q q$ meson Regge trajectories assuming
$\chi$SR.
In Sect. \ref{csvrt} we explain that linearity of the  $\bar q q$ meson trajectories
(in the absence of dislocations) implies the Nambu-Goldstone mode. 
Section \ref{prelim} presents a quasiclassical picture of high excitations of $ q\bar q$ 
mesons
which illustrates our conclusions. 
In Sect. \ref{dwisa} we consider the meson decay widths and estimate $g_A$.
Section \ref{adsqcd}
is devoted to the issue of implementation of the
chiral symmetry within the string/gauge duality formalism.
 In Sect. \ref{thera}
we discuss chiral symmetry from the Euclidean side.
Section \ref{bary} presents a brief discussion of the baryon Regge trajectories.
Finally, Sect. \ref{concl} summarizes our conclusions.
In Appendix A we review representations of the linear chiral symmetry.
In Appendix B we discuss generalized Goldberger--Treiman relations.

\section{Coexistence of the Nambu--Goldstone and
linear realizations of the chiral symmetry}
\label{coeng}

The QCD Lagrangian with $N_{f}$ massless quarks possesses U$(N_{f})_{L}\times$U$(N_{f})_{R}$
chiral symmetry (the axial U(1) is anomalous at the quantum level). 
The linear representations of this symmetry were studied in the literature in detail;
we review them in Appendix A. 

We know for certain that this chiral symmetry is spontaneously broken 
which implies the existence
of the massless Goldstone bosons, pions.
A manifestation of this breaking is non-degeneracy of the chiral partners, say,
$\rho$ and $a_1$ mesons. Thus,  at low energies the chiral symmetry is
implemented in the Nambu--Goldstone mode. 

Then, how could this symmetry be asymptotically restored in high excitations?

For illustrative purpose consider a simple case of U(1)$_{L}\times$U(1)$_{R}$
chiral symmetry,\footnote{%
\,Thus, we neglect dynamical breaking of the U(1)$_{A}$ symmetry
through anomaly.
The  effect of breaking vanishes at $N_c=\infty$. At finite $N_c$ this effect is
suppressed by assumption of asymptotic $\chi$SR.
} 
having in mind the its generalization to U(2)$_{L}\times$U(2)$_{R}$,
for two massless flavors  is quite straightforward.
In the linear realization the symmetry generators $V$ and $A$ act as
\beq
\label{genera}
V|\pm\rangle =|\pm\rangle\,,\qquad A|\pm\rangle =|\mp\rangle\,,
\eeq
where $|\pm\rangle$ are opposite parity states  whose masses are degenerate,
$M_{+}=M_{-}\,$.

The matrix element of the axial current $a^{\mu}$ between any opposite-parity 
states generically has the following form:
\beq
\label{axial}
\langle +| a^\mu |-\rangle = g(q^{2})(p_+ +p_-)_\nu\Big(g^{\mu\nu}\!-
\frac{q^\mu\,q^\nu}{q^2}\Big)=g(q^{2})\bigg[(p_+ +p_-)^\mu-q^{\mu}\,\frac{M_{+}^{2}\!-\!M_{-}^{2}}{q^{2}}\bigg],
\eeq
where  $q=p_{+}-p_{-}$ is the momentum transfer and we assume, for simplicity,
that $|\pm\rangle$ are spin-0 particles (generalization to higher spins is straightforward).
The specific form above is dictated by conservation of the axial current.  
The pole at $q^{2}=0$ reflects propagation of the massless pion.

In the linear realization, when $|+\rangle$ and $|-\rangle$ are partners,
their masses are degenerate, 
\beq
M_{+}^{2}\!=\!M_{-}^{2}
\label{threep}
\eeq
and the axial coupling 
\beq
g_{A}=g(0)=1
\label{four}
\eeq
to ensure
that the axial generator $A=\int\! {\rm d}^{3}x\, a^{0}$ acts 
as in Eq.\,\eqref{genera}. As seen from Eq.~(\ref{axial}),
the degeneracy (\ref{threep}) implies pion  decoupling.

When the chiral symmetry is spontaneously broken the masses are not equal,
$M_{+}^{2}\neq M_{-}^{2}$,  and the coupling 
to pion does not vanish, it is given by the generalized Goldberger--Treiman relation which follows from  Eq.\,\eqref{axial} (see also Appendix B),
\beq
\label{goldtrei}
g_{\pi + -}=f_{\pi}^{-1}g_{A}(M_{+}^{2}\!-\!M_{-}^{2})\,.
\eeq
There are no constraints on the axial coupling $g_{A}$ because the matrix element of the axial generator 
$A$ vanishes, $\langle -| A |+\rangle=0$, as seen from Eq.\,\eqref{axial} at $\vec q=0$.
Indeed in the spontaneously broken phase the action of $A$ just adds a soft pion to 
the $|+\rangle$ state instead of converting it into $|-\rangle$.

How is it possible to interpolate between the nonlinear realization of the chiral  symmetry 
for low-lying states and restoration of the linear realization for highly excited states?
Imagine that the mass difference $\Delta M_{\pm}=M_{+}-M_{-}$ is  small for high 
excitations, i.e., much smaller than a scale $M_{\rm h}$ at which the form 
factor $g(q^{2})$ changes. Then the matrix element \eqref{axial} for $a^{0}$ in the rest frame of $|+\rangle$ takes the form
\beq
\label{axial1}
\langle -| a^0 |+\rangle = g(q^{2})\,2M_{-}\,\frac{{\vec q}^{2}}{{\vec q}^{2}- (\Delta M_{\pm})^{2}}\,,
\eeq
where we imply $|\vec q|\ll M_{\pm}$ for the spatial momentum transfer and neglected $\Delta M/M$ 
corrections. We observe the above-mentioned  vanishing of 
this matrix element at $\vec q=0$. 

For small $\Delta M_{\pm}$ there could be a range of $\vec q$
\beq
\label{window}
\Delta M_{\pm} \ll |\vec q| \ll M_{\rm h}\,,
\eeq
where
\beq
\label{axial2}
\langle -| a^0 |+\rangle =  2M_{-}\,g(0)=2M_{-}\,g_{A}\,.
\eeq
This can be interpreted as the integral over $a^{0}$ over a finite spatial range $R\sim 1/|\vec q|$, 
in contrast with the $\vec q=0$ case when integration for the  axial charge includes all space.
Moreover, this implies that  for high excitations chiral symmetry is restored in its linear form at the scales \eqref{window} and $g_{A}=1$.

For the scenario above to work the condition $\Delta M_{\pm} \ll M_{\rm h}$ must be satisfied.
The parameter $M_{\rm h}$ is related to the splitting between 
the neighboring states of the same parity $M_{n+1}-M_{n}$ (for extended objects of size $R$ it is the same as $1/R$).
Examining Fig.\,\ref{fone}
one can see that $M_{\rm h}$ defined above is of the same order as $\Delta M_{\pm}$.
Thus,
the window \eqref{window} does not exist.

This is particularly clear for the states on the leading Regge trajectories, i.e. 
the minimal mass states for the given spin. 
There are no chiral partners on the leading  trajectory.
The splittings between the would-be 
chiral partners is of the same order as the splittings between 
the same parity states.
It shows that a simple pattern of the chiral symmetry restoration is not realized in nature,
at least in the explored range of excitations on the $\{M^2,\,J\}$ plane.

A different scenario could be realized if 
\beq
\Delta M_{\pm}^2=M_{+}^2-M_{-}^2
\eeq 
tended to zero for high excitations. Then, at sufficiently high $n$, the
splittings between the states in the would-be 
chiral pairs would become much smaller
than the  splittings between the states of the same parity
(in the U(2)$_L\times$U(2)$_R$ case, the splittings inside 
the would-be
linear chiral multiplets, see Appendix A, would become
much smaller than the splittings between
distinct multiplets). Then the pion coupling becomes weak 
and can be neglected. Simultaneously, $g_A$'s
inside the multiplet tend to 1, while $g_A$'s for transitions 
connecting distinct multiplets tend to zero. In this case
the states from the given chiral
multiplet can be considered in isolation from the rest. 
This pattern of chiral symmetry realization could be referred
to as asymptotic restoration. 

{\em A priori} the rate of vanishing of $\Delta M_{\pm}^2$
in this scenario can be arbitrary. It depends on underlying dynamics.
A natural  scaling law  is
\beq
\Delta M_{\pm}^2 \propto M^{-2}  \propto (n^{-1},\, J^{-1})\quad \mbox{or} \quad
\Delta M_{\pm}\propto M^{-3} \propto (n^{-3/2},\, J^{-3/2}) \,.
\eeq
The above expressions can serve as a benchmark scaling law.
Note that in the case of constant $\Delta M_{\pm}^2$ we get $\Delta M_{\pm}$ still decreasing for high excitations as 
$1/\sqrt{n}$. In the literature this was often viewed as sufficient for
restoration of linear chiral symmetry. As was argued above, this is not the case. The splittings of the chiral 
partners must be much smaller than other splittings to produce the 
genuine restoration.
We discuss this in more detail in the subsequent sections.

\section{Quark-antiquark meson Regge trajectories}
\label{qamrt}

Let us focus on $\bar q q$ mesons. Such states 
are characterized by the total quark spin $S=0,1$ (singlet or triplet) and
total angular momentum $J$. In nonrelativistic models one can also 
define the orbital momentum $L=J$ for singlets and $L=J\pm 1, J$ for triplets.
In the relativistic case, while $L$ itself is not a good quantum number, spatial
and charge parities (for neutral mesons),
\beq
P=(-1)^{L+1}\,, \quad C=(-1)^{L+S}\,,
\eeq
are well defined and conserved. 
Then, $L$ mod 2 and $S$ mod 2 are good quantum numbers. 
Since $S$ can be only 0 or 1 this means that $S$ is a good quantum number too.
Correspondingly, there are spin singlet Regge trajectories for which $P=-C=(-1)^{J+1}$,
and spin triplet ones with $P=C=(-1)^{J}$ and $P=C=(-1)^{J-1}$.
The parity flip (at given $J$) implies a change in $L$ by one unit.

Thus, the only new relativistic feature is a mixing between $L= J\pm 1$ states.\footnote{%
\,The importance of mixing was recently emphasized in \cite{GN}.}
There are two types of spin triplet Regge trajectories associated with two different 
combinations of $L=J+1$ and $L=J-1$.
Assuming $\chi$SR allows us two distinguish these combinations by the U(1)$_{A}$
quantum numbers.

Suppose that the chiral U$(N_{f})_{L}\times$U$(N_{f})_{R}$ would be realized linearly.
How this would be reflected in the above classification? 
In this case, as it is discussed in Appendix A,  it is convenient to classify 
the quark-antiquark pairs by their chirality content: 
$$
\bar q_{L} q_{L},\,\,\bar q_{R} q_{R}, \,\,\bar q_{R} q_{L}, \,\,\,{\rm and}  \,\,\,\bar q_{L} q_{R}\,.
$$ 
In fact, this is in one-to-one correspondence with the conserved  U(1)$_{A}$ charge which is
$\{0,\,0,\,2,\,-2\}$ for the above pairs. The first two pairs are U(1)$_{A}$ neutral while the last two are
charged. This quantum number distinguishes two types of the Regge trajectories for $S=1$
mentioned above. 

To illustrate the point let us consider some examples of interpolating quark operators, see Appendix A.
For instance, the $\rho$ meson can be represented by two such operators:
\beq
\bar q \gamma^{\mu}{\vec \tau} q\,,\qquad 
\partial_{\nu}\big[\bar q\, \sigma^{\mu\nu}\,{\vec \tau}\, q\big]\propto \bar q\, {\vec \tau}
\stackrel{\!\!\!\leftrightarrow}{D^{\,\mu}}\! q\,.
\eeq
The first one is U(1)$_{A}$ neutral, it contains a certain combination of $S$ and $D$ waves.
The second is U(1)$_{A}$ charged and pure $D$ wave. Correspondingly we have two distinct 
states with the $\rho$ meson quantum numbers  in the linear realization. What are their chiral partners? 
For the U(1)$_{A}$ neutral $\rho$ meson it is the $a_{1}$ axial meson produced by 
$\bar q \gamma^{\mu}\gamma_{5}{\vec \tau} q$. The partner to the U(1)$_{A}$ charged $\rho$ meson
is $h_{1}$ associated with the operator 
$\bar q \gamma_{5} \stackrel{\!\!\!\leftrightarrow}{D^{\,\mu}}\! q$
which is singlet with respect to both, isospin and spin. The same U(2)$_{L}\times$U(2)$_{R}$ multiplet
contains also the isotriplet axial $b_{1}$ together with isosinglet vector $\omega$. All these 8 mesons
are U(1)$_{A}$ charged. The chiral $(1,0)+(0,1)$ multiplet in the U(1)$_{A}$ neutral sector  contains 6 mesons. Besides one should also add two isosinglets, $\omega_{1}$ and $f_{1}$, which are singlets of 
U(2)$_{L}\times$U(2)$_{R}$. They are expected to be degenerate with isotriplets for dynamical reasons.
Thus, in both, U(1)$_{A}$ neutral and charged sectors we have 8 mesonic states. The same is valid for 
their Regge recurrences. 



 The U(1)$_{A}$ neutral and charged states do not mix in the linear realization.\footnote{%
 \,Although the mixing is not suppressed if chiral symmetry is spontaneously broken  the number of states (Regge trajectories) remains the same. } In each sector we have 8 mesons. In the U(1)$_{A}$ neutral sector they all have $S=1$. The mixing of $L=J\pm 1$ is fixed by U(1)$_{A}$ neutrality.
 In the U(1)$_{A}$ charged sector only $L=J + 1$ is realized (of course, $L=J$ for spin singlets
 in the multiplet).
 

If we look at the spectrum of the known mesons, see Fig.\,\ref{fone}, we do not see
the abundance of the  states expected in the linear realization.


\section{Chiral symmetry vs. linear Regge trajectories}
\label{csvrt}

In the Regge picture with the linear trajectories the $\bar q q$-meson resonances  
lie equidistantly on straight lines $M^{2}(J)=(J-J_{0})/\alpha'$ in the 
plane $\{M^{2},\,J\}$.
Distinct trajectories differ only by the intercept values $J_{0}$. 
Moreover, as seen from Fig.~\ref{fone},
on the leading trajectory, i.e., the one with no radial excitations,
 there is no  parity degeneracy in $M^{2}$. For, instance, $\rho$ lies on the leading trajectory,
while its ``partner" $a_1$ lies on the first daughter trajectory.

In general, for
 $\rho$ and $a_{1}$ trajectories,  $M^{2}$ is not degenerate,
\beq
\Delta M_{\pm}^{2}=M_{+}^{2}-M_{-}^{2}=\Delta J_{0}/\alpha'\sim \Lambda^{2}\,,
 \label{chirdep}
\eeq
i.e. $\Delta M_{\pm}^2$ does not fall off
for higher-$J$ excitations. This mass difference is of the same order as the gap between neighboring states which 
are not chiral partners, i.e. for neighboring radial excitations,
\beq
\Delta M_{\pm}^2\sim M^{2}(\rho_{n+1})- M^{2}(\rho_{n})\,,
\label{chirde}
\eeq
where the subscript $\pm$ labels the chiral partners while
$n$ refers to radial excitations.

Of course, Fig.\,\ref{fone} covers only a limited range
of $J$ and $n$. {\em A priori} one cannot rule out
that the pattern visible in this figure changes at still higher values
of $J$ or $n$. Moreover, some of the data presented
at level 5 are considered with reservations by experts.
Still, given a rather broad range of $J$ and $n$ covered, it seems natural to
think that this pattern will continue outside this range. 
Let us assume assume that:

(i) The (quark-antiquark) meson leading and daughter trajectories are linear and parallel
to each other, at least to the extent shown in Fig.~\ref{fone}
(it is natural to think that at larger values of $J$ and $n$
the degree of linearity is even higher);

(ii) These trajectories are equidistant
and there are no dislocations, i.e. each meson presented on the leading trajectory
has radial excitations at every level;

(iii) The pattern of no parity-degenerate states on the leading trajectory clearly
visible in Fig.\,\ref{fone} continues outside the range of $J$ presented in this figure.
\begin{figure}[h]
 \centerline{\includegraphics[width=5in]{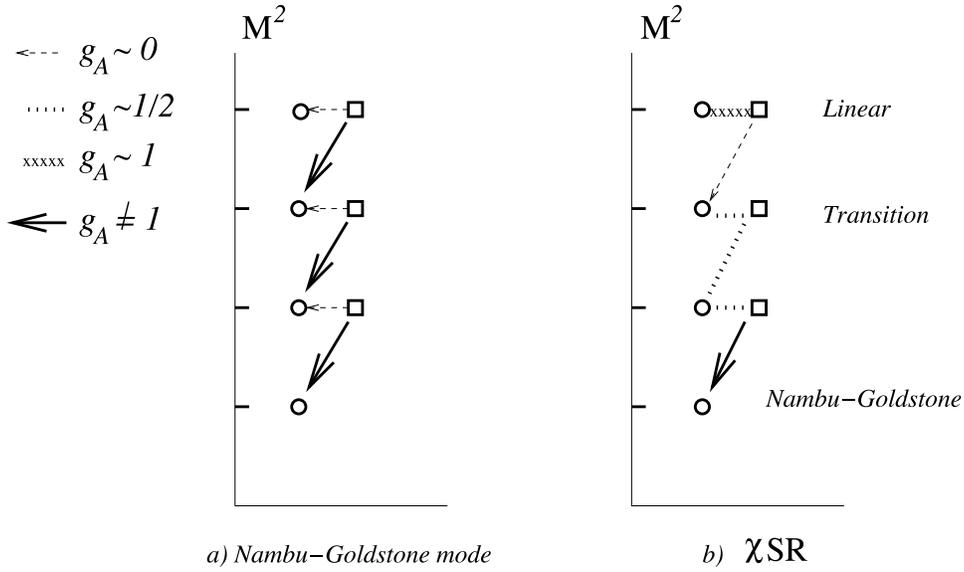}}
 \caption{\small Two scenarios of the chiral symmetry realization for high excitations.
 Open circles denote $\rho$ and its excitations, open squares $a_1$ and its excitations.
 Arrows show the values of the axial constant $g_A$ for the corresponding transitions.}
\label{twopa}
\end{figure}

\noindent
 Accepting these assumptions one cannot obtain $\chi$SR.
 
 Let us start from the $\rho$ meson. Its chiral partner is $a_1$,
the $\pi\rho a_1$ coupling is not small, and the chiral symmetry in this system is implemented in the nonlinear mode.
Moreover, there are reasons to believe that $\pi\rho a_1^\prime$ coupling is not small
either and so are  couplings between other neighboring excitations. If so, 
a few $\rho$ mesons and a few $a_1$ mesons are all entangled
in a network of chiral transitions with the pion emission.
Clear-cut  representations of U(2)$_L\times$U(2)$_R$
isolated from the rest of the spectrum have no physical grounds for existence.
Accepting assumption (i), (ii) and (iii) above
we must say that the same statement refers to high excitations as well. The value of
$\Delta M^2_{\pm}$ cannot continuously tend to zero
at large $n$, while a discontinuous jump is forbidden by assumption
(ii). No  asymptotic restoration of the chiral symmetry takes place.

This statement looks counter-intuitive.
We got used to the fact that at high energies spontaneously broken symmetries
are restored, the vacuum structure (i.e. a nonvanishing
chirally-noninvariant quark condensate) becomes unimportant.
 Equations (\ref{chirdep}) and
(\ref{chirde}) imply that \mbox{$\Delta M^2_{\pm} /M^2 \sim 1/n$}.
Although it is sufficient for symmetry restoration of inclusive quantities
(such as scalar versus pseudoscalar correlators) 
this degree of fall-off is insufficient for $\chi$SR. Note that the inclusive quantities 
correspond to measurements in the Euclidean domain, see Sect.\,\ref{thera}.

A reservation is in  order here. There is a logical possibility  of 
a more subtle ``dislocation" --- a dislocation not in the spectrum  but in the values
of $g_A$'s.
This possibility is depicted in Fig.\,\ref{twopa}.
As an example, this figure displays $\rho$, $a_1$ and their excitations.
Assume that for the lowest-lying states,
$\rho$ and $a_1$ and their close neighbors, there is 
a set of nonvanishing $g_A$'s for a number of positive-negative
parity amplitudes connecting various levels. 
The chiral symmetry is implemented in the
Nambu--Goldstone mode. For higher excitations
this pattern can either continue indefinitely (see Fig.\,\ref{twopa}\,$a$), or, after a transition domain
where $g_A$'s are reshuffled (Fig.\,\ref{twopa}\,$b$), be replaced by a ``monogamous"
behavior, with $g_A =1$ for the opposite parity states from one and the same level
and $g_A =0$ for the opposite parity states from different levels.
In this scenario with increasing $J$ the transition domain
must shift to higher levels.
We are aware of no dynamical scheme that could yield such a behavior.

The growth of characteristic distances with 
the meson mass is a natural feature
in quasiclassical string picture of hadrons. In this picture, 
to be discussed in the next section, 
high excitations correspond to extended objects with growing sizes,
while the Regge trajectories are linear.

\section{Quasiclassical Picture}
\label{prelim}

It is not clear to which extent one can 
quantify the  quasiclassical picture. Nevertheless it seems to be instructive
and it helps us
introduce the notion of a pulsating QCD string and reveal the roots of a broad
degeneracy in the meson spectrum.

Let us consider excited mesons which could be produced in $e^+e^-$
annihilation into $u$ and $d$ quarks ($\rho$'s and $\omega$'s). 
Quarks injected into the vacuum by a virtual photon at energy $E$
travel as free objects, flying back-to-back, during the time interval
$\tau_*\sim E/\Lambda^2$. During this time a string of length
$\ell_* \sim E/\Lambda^2$ develops between the endpoint quark and antiquark,
which eventually absorbs the quark kinetic energy. At distance $\ell_*$
quarks loose their kinetic energy, become nonrelativistic
(that's where $\chi$SB is crucial), turn around, move in the opposite direction, 
``head-on," and eventually stretch the string of the same length
with the positions of the endpoint
$q$ and $\bar q$ interchanged. Then the cycle repeats. In the limit $N_c\to\infty$
the string cannot be broken. Nor can it shake off a part of its energy in the
form of glueball emission.
Near each $\bar q q$ turning point all
the energy $E$ of the system resides in the stretched string.
The endpoint quarks are slow, and their chirality can (and must) be flipped.
 In this way there emerges a pulsating system of $q$ and $\bar q$ 
 connected by a long and energetic QCD string. 
The string can also rotate; its length grows with angular momentum.
 Quasiclassically, the string angular momentum and the quark spins decouple.
 (Empiric evidence of feebleness of spin interactions in high excitations
 was discussed long ago, see e.g. \cite{akai}; for a more recent consideration see
 \cite{sewi}.)

\begin{figure}[h]
 \centerline{\includegraphics[width=3in]{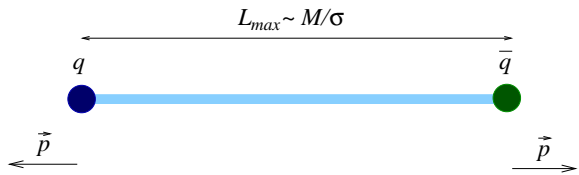}}
 \caption{\small {The quark and antiquark inside a highly excited meson,
 viewed quasiclassically,
 ``oscillate" being attached to the end-points of the string
 that does not break at $N_{c}\to\infty$. 
  }}
 \label{pk}
 \end{figure}

A snapshot of such highly excited meson in the limit $N_{c}\to\infty$
is given in Fig.\,\ref{pk}. The quark and antiquark are attached to an unbreakable string
with the tension $\sigma\sim \Lambda^2$. 
(In what follows we will omit inessential numerical constants 
and assume that the only mass dimension is provided
by $\Lambda_{\rm QCD}\equiv \Lambda$. The quark mass terms in the Lagrangian are set to zero, i.e. we will deal with the chiral limit.
In this convention the string tension is $\Lambda^2$, while the $\rho$-meson mass
is $\Lambda$.)

The mass of a high radial excitation of the meson state (say, $\rho_n$)
can be determined from a  quasiclassical 
quantization condition.   The  mass
$M_{n}$ can be presented as 
\beq
M_{n}=2p+\sigma r\,.
\eeq
The quarks  create a flux tube of the chromoelectric field
with the maximal length
\beq
\label{one}
\ell_* =\frac{M_{n}}{\sigma}\,.
\eeq
The quasiclassical quantization condition implies
\beq
\int_{0}^{\ell_*}\!\! p(r)\, dr= \pi n\,,\qquad n=1,2,\ldots
\label{two}
\eeq
with $p(r)=(M_{n}-\sigma r)/2$. Then
we immediately arrive at
\beq
M_n^2 =4\pi\sigma n \sim \Lambda^2 \, n\,.
\label{three}
\eeq
Let us parenthetically note that the asymptotically linear $n$ dependence of $M_n^2$
was analytically obtained   in the two-dimensional  't Hooft model
\cite{TM, TM1} where linear confinement is built in. In this case the next-to-leading  correction is logarithmic in $n$. 

A similar quasiclassical estimate for a spinning string implies linearity 
of $M^{2}$  in the angular momentum $L$. In fact, if both $n_r$ and $L\neq 0$,
Eq.~(\ref{three}) stays valid with the substitution
\beq
n\to n_{r}+L+1\,.
\label{sfive}
\eeq      
It is important that for high excitations
the length of the chromoelectric flux tube $ \ell$
connecting a (massless) quark $q$ with an antiquark $\bar q$
is large. It scales as $\ell \sim M.$ 

This quasiclassical string picture above refers to the spinless constituents (of the opposite parity) 
at the endpoints.
It should be supplemented by
the endpoint quark spins. For long strings we neglect spin interaction.
There are good reasons to believe 
that in this case spin interactions are weak. A theoretical argument is that 
since (chromo)magnetic charges are supposed to be condensed
in the QCD vacuum the magnetic interactions are screened.  A phenomenological argument 
is based on the pattern of degeneracy
in the observed meson spectrum \cite{akai}. The spin independence means that the spectrum is 
the same as for spinless quarks but the multiplicity is, of course, four times larger.

Consider, for instance, the $a_{1}$ meson, $J^{PC}=1^{++}$. It has $L=1$ and $S=1$.
The spin independence implies that it is degenerate with the $S=0$ state $b_{1}$.
We can show now inconsistency of the presented semiclassical picture
with the linear realization of the chiral symmetry.

Let us assume $\chi$SR.
In this case the spin degrees of freedom are untangled through the U(1)$_{A}$ classification, as was discussed in Sect.\,\ref{qamrt}. For parity partners in the chiral multiplets $L$ is shifted by one unit. This was also 
mentioned in Sect.\,\ref{qamrt}.
If the values of $n_{r}$ are the same then $M^{2}_{+}-M^{2}_{-}=4\pi \sigma$. Of course, shifting $n_{r}$
simultaneously by one unit
we could achieve the degeneracy. However, these states would not be chiral partners because their radial ``wave functions" are different and, correspondingly, $g_{A}\neq 1$.

We pause here to make an important remark.
{\sl Degeneracy of the opposite-parity states lying on the daughter trajectories
does not automatically imply chiral symmetry restoration.}
For instance, $\rho^\prime$ and $a_1$ belong to one and  the
same level of the first daughter,
but it is highly unlikely that $g_A=1$ in the $\rho^\prime\to a_1$
transition. Hence, $\rho^\prime$ and $a_1$ do not form a (part of a)
 linear representation of the chiral symmetry. Looking only at the mass spectrum
 it is easy to misinterpret such degeneracy as $\chi$SR. Note, that the small values of the 
 spectral overlap parameter, defined  as the ratio of $\Delta M_{\pm}$ to the mass gap between 
 the radial excitations, reported in \cite{Greport} are due to such choice of the would-be 
 chiral partners. For a related discussion of the spectral overlap see Ref.\cite{Cata}.

Equations  (\ref{three}) and (\ref{sfive}), taken at their face value,
lead to two crucial consequences. First, they predict the absence of degeneracy on the leading trajectory.
Indeed, for given $J$ the lowest $M^2$ state is obtained
by setting $n_r=0$ and $L=J-1$.
Equations  (\ref{three}) and (\ref{sfive}) also yield linear equidistant trajectories with degeneracies
on the daughter trajectories. For the same value of $J$ the states on the
first daughter trajectory can be obtained 
by setting $n_r=0$ and $L=J$ or $n_r=1$ and $L=J-1$ (in both cases $n$
in Eq.~(\ref{sfive}) is $J$).
Needless to say, similar pattern extends to higher daughters.

The very same behavior --- that $M^2$ depends on the combination $n_{r}+L$ 
--- was observed in the string--gauge duality analysis of long
strings in a certain approximation \cite{cobi}.
In this formalism, the string endpoint spins
are implemented by imposing specific boundary conditions, 
either of the NS or R type \cite{Vaman}.
This implies that in the approximation of
Ref. \cite{cobi1} mesons built of (hypothetical massless)
scalar quarks are degenerate with those built of the spinor quarks,
which entail the quark spin orientation independence of the meson 
mass.

It is in order to mention here  possible corrections to the above string degeneracy.
If we assume that for large $n_r,L$
\beq
M\propto \sqrt{n_r+L+1}\left[1+O({\rm max}(1/n,\,1/L))
\right]
\eeq
then deviations from the string degeneracy take the form
\beq
\delta M^2 = O({\rm max}(1/n,\,1/L)).
\eeq
This is a very interesting question which calls for further discussion.

\section{Decay widths in semiclassical approximation}
\label{dwisa}

Now let us apply a similar quasiclassical consideration to the decay widths of high radial excitations
 \cite{Nussinov,Gurvich}, see also
\cite{Gupta}. The decay 
probability (per unit time) is determined, to order $1/N_{c}\,$, 
by the probability of producing an extra quark-antiquark pair. Since 
the pair creation can happen anywhere inside the flux tube, the
probability must be proportional to $\ell$. 
As a result \cite{Nussinov,Gurvich,Gupta} one gets\,\footnote{\,This 
equation clearly demonstrates that the limits
$N_{c}\to\infty$ and $n\to\infty$ are not commutative, a rather obvious fact. 
We must first send $N_{c}$ to infinity, and only then can we consider high  
excitations.}
\beq
\Gamma_n\sim \frac{1}{N_{c}}\,  \ell_* \,\Lambda^2 =\frac{\beta}{N_{c}}\, M_n\,,
\label{estgam}
\eeq
where $\beta$ is a dimensionless coefficient of order 1 independent
of $N_c$ and $n$.

Thus, the width of the $n$-th excited state
is proportional to its mass which, in turn, 
is proportional to $\sqrt{n}$.
The square root formula for $\Gamma_n$ was numerically 
confirmed \cite{BS} in the   't Hooft model. 
It is curious that both, in actual QCD and in two
dimensions, $\beta \sim 0.5$.\,\footnote{\,Note, however, that the fit in  \cite{Afonin2} is not consistent with \eqref{estgam},
possibly due to contamination of the data set by non-$q\bar q$ mesons.}

Equation (\ref{estgam}) was recently obtained in the 
framework of string--gauge  duality in Ref.~\cite{cobi2}
which treats only the case $L\neq 0,\,\, n_r=0$.
The authors calculated a subleading correction to Eq.~(\ref{estgam})
reflecting a deviation in the linear relation $\ell_* \sim M$. Inclusion of this correction improves the fit in the low-energy domain.

The result (\ref{estgam}) can be easily understood if one takes into account
the fact that the number of open typical decay channels $\sim n$,
while each individual typical two-particle decay width is $\sim N_ c^{-1} (\Lambda^2/
M_n)$.

 \begin{figure}[h]
 \centerline{\includegraphics[width=3in]{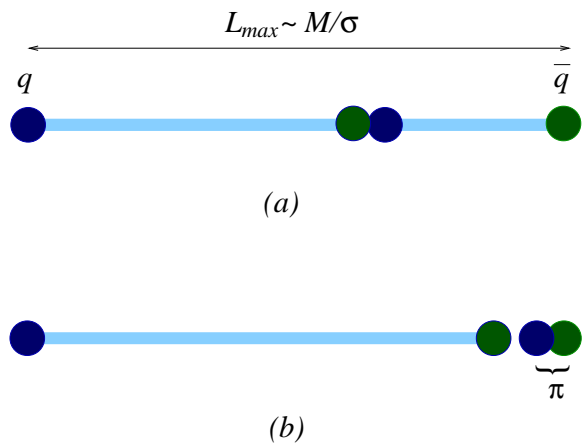}}
 \caption{\small {(a) A typical decay through pair creation
 producing two excited mesons in the final state; (b)
 A rare event with the string breaking at the end which can lead
 to pion emission. 
  }}
 \label{pkp}
 \end{figure}
 
If we are interested in the transition of the type
$A\to B+\pi$ the estimate of the decay width drastically changes. Indeed,
the pion channel is exceptional rather than typical (see Fig.~\ref{pkp}); the pion
can be produced only if the quark-antiquark pair breaks
the string of Fig.~\ref{pk} close to one of the endpoints (within distance
$\sim \Lambda^{-1}$),
rather than in the middle of the string. Then the corresponding amplitude is
\beq
\propto |\vec{p}_\pi |/\sqrt{N_c} \sim \Delta M^2 g_A /f_\pi\,.
\label{haha}
\eeq
This implies that $g_A \sim 1/\sqrt{n}$ if $A$ and $B$ are close neighbors (in mass),
and falls off fast when $A$ and $B$ become distant neighbors.
This is another argument in favor of the Nambu--Goldstone
realization of the chiral symmetry. (The second expression in Eq.~(\ref{haha})
is due to the generalized Goldberger--Treiman relation, see Appendix B.)

\section{Holographic string/gauge duals}
\label{adsqcd}

We will discuss two different approaches: top-down  
and bottom-up.
In both the holographic description of hadrons with 
the fifth coordinate  is used. 
In the first calculation by Sakai and Sugimoto  
\cite{ss} the gauge theory is represented 
by a stack of $N_{c}$ D4-branes in ten-dimensional type IIA string theory. 
One of the D4 dimensions is compactified on the circle $S^{1}$ 
to break supersymmetry of the superstring theory by anti-periodic boundary conditions for 
fermions.
The quarks are introduced by $N_{f}$ test D8-$\overline{\rm D8}$ pairs 
living in dimension orthogonal to $S^{1}$. They are associated with strings connecting
a D4 brane with D8 or $\overline{\rm D8}$; the ${\rm U}_{L}(N_{f})\times{\rm U}_{R}(N_{f})$
chiral symmetry of QCD is realized as a gauge symmetry of the $N_{f}$ D8-$\overline{\rm D8}$ pairs.

\begin{figure}[h]
 \centerline{\includegraphics[width=4in]{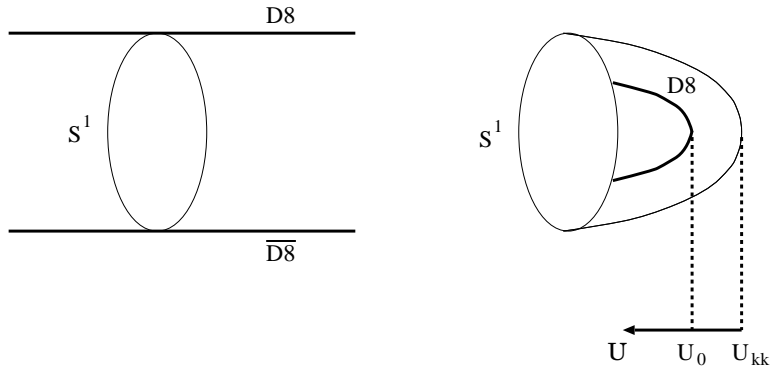}}
 \caption{\small A sketch of D8 and $\overline{\rm D8}$ branes from \cite{ss}.}
 \label{s1}
 \end{figure}

In the limit of large $N_{c}$ and large 't Hooft
coupling the supergravity approximation to string theory
is applied. The solution for the metric is characterized by existence of the horizon, $U>U_{\rm KK}$,
in the radial coordinate $U$ transverse to the D4 branes. As $U\to U_{\rm KK}$ the radius of $S^{1}$
shrinks to zero and D8/$\overline{\rm D8}$ branes merge to form a single component of the D8-branes,
see Fig.\,\ref {s1}. 
Only the diagonal ${\rm U}(N_{f})$ survives on the resulting D8 brane. This explicitly shows 
 spontaneous breaking of the ${\rm U}_{L}(N_{f})\times{\rm U}_{R}(N_{f})$ chiral symmetry.

In application to the vector, axial-vector and spin-zero fields the above construction
generates the action
\beq
S^{\rm SS} =\kappa\,\int\, d^4x\,dz\, {\rm Tr}\,
\left[K^{-1/3} \, \frac{1}{2}\, F_{\mu\nu}^2 + K\, F_{\mu z}^2
\right]
\eeq
where
\beq
K=1+z^2\,,\qquad \kappa =\frac{\lambda\,N_c}{108\pi^3}\,.
\eeq
This expression is written in five dimensions, $z$ is the holographic coordinate.
Boundaries at positive and negative $z$-asymptotics correspond to $V\pm A$ combinations
of the gauge fields while $A_{z}$ is associated with the pseudoscalar field.  The conformal symmetry
is softly broken; one-dimensional quantum mechanics in the holographic direction gives the mass spectrum. The massless pion is built-in and does not decouple from high excitations.
Related to this is a nice feature of the construction: a Skyrmion nature of baryons.
What is absent in this picture is the linearity of the Regge trajectories; they are parabolic instead.

In the second, bottom-up approach, the authors  introduce 
independent fields in each channel, i.e. vector, axial-vector and spin-zero fields.
They also fix the 
metric and dilaton field to get linear dependence of $M^2$
on  $n_r$    and  $L$,
\beq
S^{\rm KKSS} =\kappa\,\int\, d^4x\,dz\,e^{\Phi (z)}\,\sqrt{g}
\left[-|DX|^2 + 3|X|^2-\frac{1}{4g_5^2}\, \left(F_L^2+F_R^2
\right)
\right]
\eeq
where
\beq
\Phi = z^2\,,\qquad g_5^2 =\frac{12\pi^2}{N_c}\,,
\eeq
while the metric is given by
\beq 
ds^{2}=\frac{1}{z^{2}}\big(\eta_{\mu\nu}dx^{\mu}dx^{\nu}+dz^{2}\big).
\eeq
As in the previous case the quantum-mechanics eigenvalues along the holographic coordinate $z$
give the meson spectrum $M^2$ in four dimensions. High excitations are  characterized by the growing
sizes both in 4d and in the holographic coordinate, $z\sim \sqrt{n}$.  

\begin{figure}[h]
 \centerline{\includegraphics[width=3in]{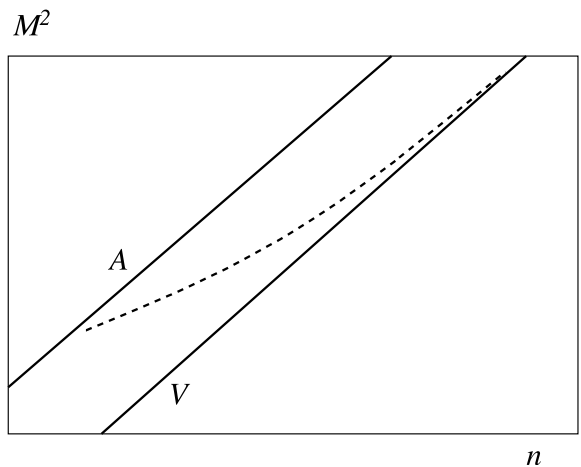}}
 \caption{\small $M^{2}$ versus $n$ for the vector and axial vector resonances from KKSS.}
 \label{adpict}
 \end{figure}

The linear chiral symmetry implies that $\langle X \rangle =0$.  Then $\rho$ and $a_{1}$ mesons are
degenerate together with the tower of their excitations. 
The splitting between the $\rho$ and $a_1$ which certainly exists in nature is
due to condensation of  the $X$ field containing pion and its scalar partner, $\langle X \rangle \neq 0$.
In the KKSS approach this splitting is proportional to 
$X^{2}/z^{2}$. Since the characteristic $z^{2}\sim n$ for high radial excitations 
the mass splitting (and corresponding symmetry restoration) depends on behavior 
of $X$ at large $z$. KKSS assumed that $X\to {\rm const}$ which means that 
 the $X$ contribution diminishes with $n$ implying
asymptotic restoration of the chiral symmetry. Clearly it simultaneously introduces
nonlinearity of the $n$ dependence of $M^{2}$, as illustrated in Fig.\,\ref{adpict}.
The solid lines in this figure depict trajectories for $\langle X\rangle=0$
while the dashed line represent the $1/n$ deformation of the axial masses due to
$\langle X\rangle={\rm const}$.
This does not seem to be compatible with 
phenomenology of known resonances which form linear trajectories starting
from  $n\sim 1$.

Moreover, there is no theoretical justification for the above assumption $X\to {\rm const}$:
despite the fact the dimension for the $X$ field is $(-3)$ there is no physical instability.
In the recent papers\cite{CKP,Bergman} the authors argued that
instead one can take  $X\sim z$ at large $z$. Then
the $M^{2}$ splittings between the chiral partners do not depend on $n$ which fits well 
with the linearity and equidistance of the Regge trajectories.

\section{Chiral symmetry in the Euclidean domain and the meson spectrum}
\label{thera}

Operator product expansion (OPE) shows that the chiral symmetry is restored at short Euclidean 
distances. For instance, for the difference of two-point functions of scalar and pseudoscalar currents
we have the following 
 OPE expansion at large Euclidean momentum $Q$
 \beq
 \Pi_S(Q) - \Pi_P(Q)\sim \frac{g^2\langle\bar q q\rangle^2}{Q^4} + ... 
 \label{ope}
 \eeq
 as was shown long ago in \cite{SVZ}. Here 
 \beqn
&& \Pi_{S,P}(Q) =-i \int d^{4}x \,{\rm e}^{iqx} {\rm T}\langle j_{S,P}(x) j_{S,P}(0)\rangle\,,\nonumber\\
 &&j_{S}=\bar q q\,,\quad j_{P}=i\bar q \gamma_{5}q\,.
 \label{ope1}
 \eeqn
 We see that the $\Pi_S - \Pi_P$ difference is expressed in terms of the quark condensate
 $\langle \bar q q\rangle$ which is the order parameter for the chiral symmetry breaking.
 We observe a rapid chiral symmetry restoration at large $Q$. 
 
 The question we address is what constraints on the spectrum follow from this behavior. 
 Of course, for any given spectral model Eq.\,\eqref{ope} does lead to constraints. For instance,
 for equidistant spectra, $M_{n}^{2} \propto n$, and equal residues one can rule out different slopes
 (see Ref.\cite{CKP} for a noncomplying example). 
  
  However, the Euclidean behavior is sensitive strictly speaking only to averaged features 
  of the spectral densities. For instance, one can replace the sum of the narrow peaks by 
  a smooth curve (quark-antiquark continuum) introducing only an exponential correction 
  of the type $\exp(-Q)$ not visible in OPE.\footnote{%
  \,Let us remind that this circumstance was used by Migdal\cite{Migdal} long ago.}
  Thus, there is no unique prediction for the mass splitting of the chiral partners,
as was emphasized, in particular,  in Refs.\cite{Cata,GM}. 

  Let us consider some limiting cases.  The first scenario is given by an appropriate splitting of the lowest resonances in the scalar and pseudoscalar channels,
 \beq
 \Pi_S(Q) - \Pi_P(Q)=F^{2}\left(\frac{1}{Q^{2}+M_{S}^{2}}-\frac{1}{Q^{2}+M_{P}^{2}}\right)
 \to -\frac{F^{2}(M_{S}^{2}-M_{P}^{2})}{Q^{4}}\,.
  \label{ope2}
 \eeq
Here we assumed that the residues are the same to eliminate the unwanted $1/Q^{2}$
term. In addition,
we assumed all higher resonances to be fully degenerate, both in masses and residues. 

However, the full degeneracy 
 for higher states can be replaced by a much weaker requirement of a local conspiracy:
the scalar and pseudoscalar resonances can be split into local in energy groups such 
that their contributions into the dispersion integrals are maintained the same.

The simple illustration of how this might work is as follows.
Consider a pair with some large $M$ in Eq.\,\eqref{ope2}. This pair contributes
$F^{2}\Delta M^{2}$ to the coefficient of $1/Q^{4}$. The residues grow with $M$ as $M^{2}$
because in 
the continuum (quark) spectral densities ${\rm Im} \Pi_{S,P}(s)\sim s$.
This implies that $M^{2}\Delta M^{2}\sim {\rm const}$ for high excitations.
This estimate was obtained in Ref.\cite{2}.\footnote{%
\,Note that a similar estimate gives the correct mass splitting in the 2d 't Hooft model.
In this case $F^{2}$ does not grow with $M$ implying $\Delta M^{2}\sim {\rm const}$.}
 Clearly, the group of conspirators can be enlarged.
If we included two scalars and two pseudoscalars in such a group we could arrange for cancellation of
$1/Q^{4}$ terms without requiring $\Delta M^{2}\to 0$. Involving more and more conspirators we could achieve exponential accuracy mentioned above\cite{Migdal}.

Note that even if $\Delta M^{2}\to 0$ at large $M$, it is not sufficient to conclude that this pair
belongs to the same linear chiral multiplet.

\section{Baryons}
\label{bary}

The problem of $\chi$SR can be addressed in baryons, with reservations,
only at intermediate values of $J$ and $n$. This is due to the fact that
baryon masses grow with $N_{c}$ and  their decay widths, generally speaking, do not
vanish in the limit $N_{c}\to\infty$. We will discuss $\chi$SR in baryons with due caution.

There are a few folklore statements regarding baryons
which do not seem to fall in one and the same picture.
Let us briefly review them.

It is believed that the nucleon and $\Delta$ Regge trajectories are linear
up to $J=9/2$, and even higher in certain instances \cite {Collins}. 
It is firmly established that for the ground states, $N$ and $\Delta$,
there are no degenerate parity partners in the spectrum. At the same time,
Particle Data Group reports degenerate $I(J^P)=\frac{1}{2}\left(\frac{5}{2}^\pm
\right)$ states on the leading trajectory. With less certainty one can speak of
parity degeneracy on the leading trajectory at $J=9/2$. 

What kind of degeneracy should one expect if  $\chi$SR takes place?
Instead of the quark-antiquark pairs  in the meson case,  for baryons we should consider quark triplets
\beq
q_{L} q_{L}q_{L}\,,\; q_{L} q_{L}q_{R}\,,\; q_{L} q_{R}q_{R}\,,\; q_{R} q_{R}q_{R}\,.
\eeq
The U(1)$_{A}$ charges of these triplets are $\{ 3\,, 1\,, -1\,, -3\}$. For the total quark spin $S=3/2$
we have two types of the chiral multiplets differing in their U(1)$_{A}$ charges:
one with the U(1)$_{A}$ charge 3, and one with with the U(1)$_{A}$ charge 1.
For $S=1/2$ there
are three types: one with the U(1)$_{A}$ charge 3, and two with with the U(1)$_{A}$ charge 1.
As in mesons, long strings imply spin degeneracy, i.e. degeneracy between 
distinct U(1)$_{A}$ charges.
Such high level of degeneracy is not yet observed in the baryon spectrum.

From the Regge theory side
 parity degeneracy in baryons  follows from the Gribov--MacDowell symmetry and linearity
of baryonic Regge trajectories \cite{g1,g2,Collins}. The linearity of the  nucleon
trajectory is supported also by data at negative $t$, in the scattering region.
The absence of the parity doublers for
$N$ and $\Delta$ and their apparent presence at $J=5/2$
could be reconciled  with the linearity of the
$P=-1$  Regge trajectory if the residues of the lowest states vanish \cite{akai}. 
This is a rather contrived scenario, though.
If we do not want such a contrived scenario then the absence of the Gribov--MacDowell parity doubling for $N$ and $\Delta$
could signal  that (some of) the baryon trajectories are nonlinear to a significant extent.
Then  a direct experimental confirmation
of this nonlinearity is a must.

As was noted in \cite{akai} (see also \cite{old} for an earlier discussion),
the meson and baryon Regge trajectories are similar in many respects,
in particular, the slopes are practically equal. This might mean that
for high excitations, when the connecting string is  long, the difference
between mesons and baryons is only at the endpoints:
quark and antiquark in the former case and quark and diquark in the latter
\cite{old,akai,sewi}.\footnote{Additional
arguments on diquarks can be found in \cite{shiva}.}
 If this is the case, the emergence of the 
parity doubling at large $J$ could be a natural consequence of the fact
that ``good" diquarks (i.e.
those with the vanishing spin and isospin) can be both
scalars (i.e. $P=+1$) and pseudoscalars (i.e. $P=-1$).
For instance, in the instanton liquid model the scalar diquark structure
is very similar to that of pions \cite{Shur}, while pseudoscalar diquarks 
are similar to $\sigma$. Due to the strongest attraction, the scalar diquarks
are the lightest, with an effective mass around 200 MeV \cite{Shur,sewi}.
In the pseudoscalar diquarks the attraction is weaker, but if they are indeed 
similar to $\sigma$ their mass can be as low as
$\sim 400$ MeV \cite{HL}. Since in high excitations the diquarks are ultrarelativistic,
their mass $m$ enters in the baryon mass in the form $m^2/p_{\rm char}$,
where $p_{\rm char}$ is a characteristic diquark momentum.

At small $J$, when there are no long strings,
the quark-diquark configuration is not necessarily dominant
over three-quark configurations.
This would naturally explain a curvature
in the $P=-1$ baryon trajectory at small $J$ and the emergence of degeneracy with
the $P=+1$ baryon trajectory at large $J$.

If this explanation is correct the corresponding parity degeneracy has nothing to do with 
$\chi$SR. Indeed, in the case of the linear realization 
of the chiral symmetry,
the U(1)$_{A}$ charge is
conserved. Then, similar to the mesonic case, the opposite parity states in the chiral multiplets are due to the shift of the 
orbital momentum by one unit rather than 
due to the passage from  the scalar diquark to pseudoscalar one or vice versa.
Just as in the meson case, this situation implies incompatibility of the $\chi$SR with the semiclassical limit.

A remark is in order here concerning a recent analysis \cite{Gloz5} of the
pion coupling to (excited) nucleons. A significant suppression
of the pion coupling was found in a number of decays of excited 
nucleons into $N\pi$
implying, through the generalized Goldberger--Treiman relation,
a suppression in the corresponding axial couplings.
We would like to point out that this observation is insufficient
for the conclusion of $\chi$SR (although, it is necessary, of course).

Indeed, in the linearly realized chiral symmetry
all transitions inside chiral multiplets have $g_A=1$ (up to Clebsch-Gordan coefficients),
while in the transitions leading outside the given chiral
multiplet $g_A=0$. In the Nambu--Goldstone mode
the values of $g_A$ for all transitions can be arbitrary.  In particular,
in the pion decays of highly excited states (long strings)
it is natural to expect that $g_A$'s are suppressed, see Eq.~(\ref{haha})
and the subsequent discussion. To demonstrate that
$\chi$SR takes place one needs to identify degenerate chiral partners
and demonstrate that $g_A=1$ for transitions inside the chiral multiplet.

\section{Conclusions}
\label{concl}

This article grew as a continuation of the ongoing heated
debate in the literature regarding asymptotic symmetries of the meson spectrum in
QCD \cite{Gloz2,Gloz0,Gloz1,Gloz3,Gloz4,2,Jaffe,cg,Cata,Afonin2},
and numerous discussions of this issue at various conferences. At an early stage
we believed that $\chi$SR could be natural in QCD. Further 
more careful studies  made us change our minds. 

Our analysis consists of two parts:
the first one is based on the existing Regge phenomenology;
in the second part we try to combine various theoretical arguments, such as quasiclassical considerations, AdS/QCD, and so on. 
All arguments are consistent with the absence of the chiral symmetry restoration
in the {\em observed} meson spectrum.

Given the abundant Regge phenomenology
plus plausible theoretical arguments, we believe that
the chiral symmetry is not  restored in highly excited mesons.
 Various arguments show  that $\Delta M^{2}$ for would-be chiral partners 
 does not depend on $n$. It means that $\Delta M\sim 1/M$ and decreases for high excitations
 as $1/\sqrt{n}$.
This is insufficient for chiral symmetry restoration.
The restoration takes place only when the chiral splitting 
 is much smaller than the splitting between, say, the neighboring radial excitations.
 In actuality
 the values of $\Delta M_{\pm}$ in the would-be chiral multiplet
are of the same order of magnitude as the splittings
of the mesons lying on distinct daughter trajectories.
There are also no obvious reasons for
$g_A $ to approach unity in the transitions
between the  chiral partners, moreover, we argued that it does not.

Our consideration is not a theorem but, rather,
a physical argument. In the present-day theory it seems
impossible to prove a theorem of no $\chi$SR
at large $n,J$, beyond the explored domain of resonances.
If such  $\chi$SR does take place, 
the quasiclassical picture must badly fail,
for reasons of which we have no clue.
The Regge trajectories
{\em must} curve in an ``intermediate window" 
which, by itself, would present a very remarkable phenomenon.

The issue of the microscopic realization of
the Gribov--MacDowell symmetry in baryons is contentious. Perhaps, the baryon data hint that linearity must be abandoned
at least for some trajectories.
For a breakthrough, 
a fully developed theory which would combine a picture of long strings for high excitations
with the chiral symmetry of massless quarks at the endpoints
is badly needed. 

\section*{Acknowledgments}

We are deeply indebted to L.~Glozman and A. Kaidalov for extensive correspondence 
and patient explanations regarding the spectrum of excited 
meson and baryon resonances, for sharing with us their insights and knowledge.
We would like to thank R.~Jaffe whose remarks and private communications
served as a decisive impetus for the completion of this work.
We are thankful to S.~Afonin, A.~Andrianov, A.~Armoni, O.~Cato, M.~Chizhov, T.~Cohen, A.~Gorsky,
T.~Gherghetta,  L.~Glozman, M.~Golterman, A.~Kaidalov, A.~Parnachev, S.~Peris,  K.~Peeters, A.~Polyakov, A.~Selem, E.~Shuryak, J.~Sonnenschein, 
M.~Stephanov, M.~Strassler, D.~Vaman, and M.~Zamaklar for very useful discussions
and informative communications.
We would like to thank the Benasque Center for Science where a  part of the work  was
carried out  in the course of the Workshop ``QCD and String Theory," July 2--14, 2006, Benasque, 
Spain, and the Isaac Newton Institute for Mathematical Sciences
where a  part of the work  was
carried out  in the course of the program ``Strong Fields, Integrability and Strings,''
23 July -- 21 December 2007.

This work is
supported in part by DOE grant DE-FG02-94ER408.

\newpage
\section*{Appendix A}

\renewcommand{\theequation}{A.\arabic{equation}}

\setcounter{equation}{0}

\renewcommand{\thesubsection}{A.\arabic{subsection}}
\setcounter{subsection}{0}

In this Appendix we review
some material relevant to the linear realization of the chiral symmetry.

If both left- and right-handed  SU($N_{f}$)'s are linearly realized, hadronic states 
must fall into degenerate multiplets of the full chiral symmetry.
The degeneracy is lifted by an ${\rm SU}(N_{f})_{L}\times {\rm SU}(N_{f})_{R}\to {\rm SU}(N_{f})_{V}$ 
spontaneous breaking of the symmetry.
Let us first briefly review an appropriate representation theory both for mesonic and baryonic states.
These representations were studied long ago, even  before the advent of QCD \cite{jido}.
We would not touch the case of heavy-light mesons which was also studied \cite{BEH}.

We will present the construction using interpolating  meson and baryon currents written
in terms of quark fields. While this is done for an illustrative purpose some features like
operator twists could be related with the stringy nature of the excited states.

\subsection{Chiral symmetry: linear realization for mesons}
\label{chira}

In terms of the quark fields the ${\rm SU}(N_{f})_{L}\times {\rm SU}(N_{f})_{R}$
chiral symmetry is conveniently represented in terms left- and right-handed 
Weyl spinors, $q_{L,R}=(1\mp\gamma_{5})q/2$,
\beq
[q_{L}]_{\alpha}^{i f}\,,\qquad [q_{R}]_{\dot\alpha}^{i\bar f}\,,
\eeq
where $\alpha,\, \dot\alpha=1,2$ are spinorial indices of the Lorentz group,
$i=1,\ldots,N_{c}$ is the color index and $f,\,\bar f=1,\ldots,N_{f}$
are subflavor indices of two independent, left and right, SU($N_{f}$).
The chiral symmetry transformations are
\beq 
q_{L}^{f}\to L^{f}_{g}\,q_{L}^{g}\,, \qquad q_{R}^{f}\to R^{\bar f}_{\bar g}\,q_{R}^{\bar g}\,,
\eeq
where $L$ and $R$ are the SU$(N_{f})_{L,R}$ matrices. Classically QCD has also U(1)$_{L}\times$U(1)$_{R}$
invariance, 
\beq 
q_{L}\to {\rm e}^{i\eta_{L}} \,q_{L}\,, \qquad q_{R}\to {\rm e}^{i\eta_{R}}\,q_{R}\,,
\eeq
The diagonal U(1)$_{V}$ (when $\eta_{L}=\eta_{R}=\eta_{B}/N_{c}$) is associated with the baryon
charge while the axial U(1)$_{A}$ (when $\eta_{L}=-\eta_{R}=\eta_{A}$) becomes anomalous at the quantum level. This dynamical breaking of U(1)$_{A}$ is suppressed at large $N_{c}$ together
with quark-gluon mixing. It also dies out for high excitations corresponding to short distances 
so we deal there with the asymptotic ${\rm U}(N_{f})_{L}\times {\rm U}(N_{f})_{R}$ symmetry.

The interpolating fields for colorless hadrons can be simply constructed 
from quark fields. Let us consider spin zero mesons. They are described by
the matrix $M$,
\beq 
\label{Mmes}
M^{f}_{\bar f} =[\bar q_{R}]^{\alpha}_{i \bar f}[q_{L}]_{\alpha}^{i f}=
\bar q_{\bar f}\,\frac{1-\gamma_{5}}{2}\,q^{f}\,.
\eeq
The baryon charge of $M$ clearly vanishes
while the U(1)$_{A}$ charge is 2.
The meson matrix $M$ realizes the $\{N_{f}, N_{f}\}$ representation of  ${\rm SU}(N_{f})_{L}\times {\rm SU}(N_{f})_{R}$  and 
contains $2N_{f}^{2}$ real fields. The reflection of space coordinates,
$P$, which transforms $q_{L\,\alpha}^{\,\,if}$ to $q_{R\,\dot\alpha}^{\,\,i\bar f}$ and vice versa, acts on the matrix $M^{f}_{\bar f}$ as
\beq
P M=M^{\dagger}\,.
\eeq
It means that the Hermitian part  of $M$ describes $N_{f}^{2}$ scalars
while anti-Hermitian part represents $N_{f}^{2}$ pseudoscalars.
In terms of the diagonal SU($N_{f})_{V}$ symmetry (when $L=R$) these $N_{f}^{2}$  fields
form the adjoint representation and the singlet. 

This construction can be easily generalized to include higher spins, the Regge recurrences 
of spin zero,
\beq 
\label{recur}
\left[M_{\mu_{1}\ldots\mu_{n}}\right]^{f}_{\bar f} =[\bar q_{R}]^{\alpha}_{i \bar f}\stackrel{\,\leftrightarrow}{D}_{\mu_{1}}\ldots \stackrel{\,\leftrightarrow}{D}_{\mu_{n}}[q_{L}]_{\alpha}^{i f}\,.
\eeq
All these operators have leading twist 3.

Starting from spin 1 there exist interpolating $q\bar q$ operators of a different chiral structure.
In case of spin 1 mesons one can introduce 
\beq
\label{sp1}
\Big [V^{L}_{\mu}\Big]^{f}_{g}=\sigma_{\mu}^{\alpha \dot\alpha}[\bar q_{L}]_{\dot\alpha\,i g}\,[q_{L}]_{\alpha}^{i f}
=\bar q_{g}\gamma_{\mu}\,\frac{1-\gamma_{5}}{2}\,q^{f}\,,
\eeq
where $\sigma^{\mu}=\{1,\vec \sigma\}$. 
The correspomding baryon and U(1)$_{A}$ charges are zero.
Subtracting the trace we get the $\{N_{f}^{2}-1, 1\}$ representation
while the trace part is the $\{1,1\}$ representation of ${\rm SU}(N_{f})_{L}\times {\rm SU}(N_{f})_{R}$.
The matrix $V_{\mu}^{L}$ is Hermitian so it represent $N_{f}^{2}$ fields of spin 1.
These fields are singlets of SU($N_{f})_{R}$ and  adjoints or singlets 
of SU($N_{f})_{L}$ (as well as SU($N_{f})_{V}$).  
Under the parity transformation $V_{\mu}^{L}$
goes to 
\beq
\label{sp11}
\Big [V^{R}_{\mu}\Big]^{\bar f}_{\bar g}=\sigma_{\mu}^{\alpha \dot\alpha}[\bar q_{R}]_{\alpha\,i\bar g}[\,
q_{R}]_{\dot\alpha}^{i \bar f}
=\bar q_{\bar g}\gamma_{\mu}\,\frac{1+\gamma_{5}}{2}\,q^{\bar f}\,.
\eeq
The vector and axial-vector particles are described by a sum and difference of 
$V_{\mu}^{L}$ and $V_{\mu}^{R}$. The Regge recurrences are obtained in the same as in Eq.\,\eqref{recur}. All these operators have twist 2. 

As was discussed in the literature, spin 1 mesons can be also described  by antisymmetric tensor field
which is the  $(0,1)+(1,0)$ representation of Lorentz group instead of  $(1/2,1/2)$ used above, 
\beq
\Big[H_{\mu\nu}\Big]^{f}_{\bar f}=[\sigma_{\mu}\bar\sigma_{\nu}]^{\alpha\beta}
\Big\{[\bar q_{R}]_{\alpha\,i \bar f}\,[q_{L}]_{\beta}^{i f}+\alpha \leftrightarrow \beta\Big\}=
\bar q_{\bar f}\sigma_{\mu\nu}\,\frac{1-\gamma_{5}}{2}\,q^{f}\,,
\eeq
where  $\bar\sigma^{\mu}=\{1,-\vec \sigma\}$.
The chiral features of this tensor current are different from $ [V^{L}_{\mu}]^{f}_{g}$ but the  same as those 
of spin 0 fields $M^{f}_{\bar f}$, Eq.\,\eqref{Mmes}, and their Regge recurrences $[M_{\mu_{1}\ldots\mu_{n}}]^{f}_{\bar f} $, Eq.\,\eqref{recur}. Moreover, by applying the  total derivative
we see that the tensor current $[H_{\mu\nu}]^{f}_{\bar f}$ is equivalent to $[M_{\mu}]^{f}_{\bar f} $.
Indeed,
\beq
\partial^{\nu}\big[H_{\mu\nu}\big]^{f}_{\bar f}=
-i\bar q_{\bar f}\stackrel{\leftrightarrow}{D_{\mu}}\,\frac{1-\gamma_{5}}{2}\,q^{f}\,.
\eeq

Thus, for any given spin we have two types of the chiral multiplets:  charged 
and neutral with respect to U(1)$_{A}$.  Each multiplet contains $2N_{f}^{2}$ states.
This accounts for degeneracy of flavor adjoints and singlets in the large $N_{c}$ limit
in case of the U(1)$_{A}$ neutral interpolating currents, as in Eqs.\,(\ref{sp1},\,\ref{sp11}).
Each multiplet contains $N_{f}^{2}$ states of each parity. For the U(1)$_{A}$ neutral 
multiplets $CP=1$ (for electrically neutral states) while in the U(1)$_{A}$ charged 
multiplets $CP=-1$. 

Spin zero is special: only higher twist U(1)$_{A}$ neutral operators
of the type $\bar q_{g}\,\gamma^{\mu}(1-\gamma_{5})G_{\mu\nu}D^{\nu}q^{f}$
 are possible. These operators correspond to hybrid mesons rather than to quark-antiquark ones.

We pause here to make two remarks
about  the $N_{f}=2$ case. Due to
quasireality of the fundamental representation of  
SU(2) the eight-dimensional representation of \mbox{${\rm SU}(2)_{L}\!\times {\rm SU}(2)_{R}$} given by 2$\times$2 matrix $M^{f}_{\bar f}$ becomes reducible and can be split into two four-dimensional ones. 
This can be done by imposing the  group-invariant conditions,
\beq
\tau_{2}M^{*}_{\pm}\tau_{2}=\pm M_{\pm}\,.
\eeq
Then 
\beq
\label{u2u2}
M_{+}=\sigma -i\,\bm{\tau \pi}\,, \qquad M_{-}=i\eta +\bm{\tau \sigma}\,,
\eeq
where all fields are real.
The quadruplet $M_{+}$ contains the isosinglet scalar $\sigma$ and the isotriplet 
of pseudoscalars {$\bm\pi$} while in $M_{-}$ the pseudoscalar $\eta$ is isosinglet 
and scalars form the isotriplet  {$\bm\sigma$}.

However, as we mentioned above the asymptotic symmetry includes also U(1)$_{A}$
(the vector U(1)$_{B}$ does not act on mesons).  The U(1)$_{A}$ transformations mix 
$M_{+}$ and $M_{-}$ thus restoring eight-dimensional representation.

The second remark refers to a hypothetical QCD-like theory
(which may or may not be useful, say, in technicolor)
rather than to actual QCD. Assume that we consider an SU($N_c)$
Yang--Mills theory with two Dirac quarks in the
adjoint representation of SU($N_c)$, or, which is the same,
four Weyl spinors in the adjoint. 
In this case, as well-known \cite{dp}, the pattern of the $\chi$SB
is different from that in conventional QCD, namely, SU(4)$\to$O(4).
Since the O(4) symmetry which is isomorphic to SU(2)$\times$SU(2)
is strictly unbroken, all hadrons in this theory, including pions and 
other low-lying states,
will be classified in multiplets of the exact SU(2)$\times$SU(2) symmetry. 

\subsection{Chiral symmetry: linear realization for baryons}
\label{chiraB}

The baryon currents can be introduced in a similar way. They contains $N_{c}$ quark fields so, for example, the baryon current with the  maximal spin $N_{c}$/2 is
\beq
\big[ B_{\alpha_{1}\ldots\alpha_{N_{c}}}^{L}\big]^{f_{1}\ldots f_{N_{c}}}=
\epsilon_{i_{1}\ldots i_{N_{c}}}q_{L\alpha_{1}}^{i_{1}f_{1}}\ldots\, q_{L\alpha_{N_{c}}}^{i_{N_{c}}f_{N_{c}}}\,,
\eeq
where symmetrization over $\alpha_{1}\ldots\alpha_{N_{c}}$ as well as over 
$f_{1}\ldots f_{N_{c}}$ is implied.
Because of symmetry in $f_{1}\ldots f_{N_{c}}$ the number of fields in the multiplet
is equal to the binomial coefficient $C(N_{c}+N_{f}-1, N_{c})$. 
Their baryon charge is 1 and the U(1)$_{A}$ charge is $N_{c}$.
The parity transformation relates 
$ \big[B_{\alpha_{1}\ldots\alpha_{N_{c}}}^{L}\big]^{f_{1}\ldots f_{N_{c}}}$ to a similarly defined
$ \big[B_{\dot\alpha_{1}\ldots\dot\alpha_{N_{c}}}^{R}\big]^{\bar f_{1}\ldots \bar f_{N_{c}}}$.

This $B_{L}^{f_{1}\ldots f_{N_{c}}}$, $B_{R}^{\bar f_{1}\ldots \bar f_{N_{c}}}$ construction for the baryons 
does {\em not} allow one to introduce the chirally invariant mass, in contrast to the meson case.
Indeed, the Lorentz invariant convolution  $\bar B_{R\alpha_{1}\ldots\alpha_{N_{c}}}^{\bar f_{1}\ldots \bar f_{N_{c}}}B_{L\beta_{1}\ldots\beta_{N_{c}}}^{f_{1}\ldots f_{N_{c}}}\epsilon^{\alpha_{1}\beta_{1}}
\ldots \epsilon^{\alpha_{N_{c}}\beta_{N_{c}}}$ contains no chiral singlet. To allow for the invariant mass one needs to add 
$\tilde B_{L}^{\bar f_{1}\ldots \bar f_{N_{c}}}$, $\tilde B_{R}^{f_{1}\ldots f_{N_{c}}}$: this is called mirror doubling \cite{jido}.
These mirror baryons cannot be introduced just as a simple product of quark fields
for which handedness of their  Lorentz index  defines handedness of of their chiral representation.
One has to use objects such as covariant derivatives $D_{\alpha\dot\alpha}=(\sigma^{\mu})_{\alpha\dot\alpha}D_{\mu}$ for the construction. This considerably increases the twist, while $t=N_{c}$ for 
$B_{\alpha_{1}\ldots\alpha_{N_{c}}}^{L}$ it twice as large, $t=2N_{c}$, for the mirror current.
This high twist could be an extra argument against asymptotic linear symmetry.

Note that the chiral invariant baryon mass requires mirror doubling for other spin and flavor assignments as well.

\subsection{Hadronic couplings: \boldmath{U(2)$_{L}\times$U(2)$_{R}$} without spontaneous breaking}
\label{hadro}

Let us consider couplings of spin 0 and 1 mesons to spin 1/2 baryons in the linearly realized {U(2)$_{L}\times$U(2)$_{R}$ as an illustrative example. As we discussed above eight spin 0 mesons ($\sigma,\,\bm{\sigma}, \,\eta,\,{\bm \pi}$)
 are described by the matrix $M^{f}_{\bar f}$, the spin 1 isotriplet mesons (${\bm \rho},\,{\bm a_{1}}$)
are given by the traceless $[V_{\mu}^{L}]^{f}_{g}$ and $[V_{\mu}^{R}]^{\bar f}_{\bar g}$
and baryons are presented by $B^{L\,f}_{\alpha},\,B^{R\,\bar f}_{\dot\alpha}$ and 
$\tilde B^{L\,\bar f}_{\alpha},\,\tilde B^{R\,f}_{\dot\alpha}$.

Free baryons are described by Lagrangian
\beq
\begin{split}
{\cal L}_{B}=& ~i\bar B_{L\,f}\gamma^{\mu}\partial_{\mu}B_{L}^{f}
+i\bar B_{R\,\bar f}\gamma^{\mu}\partial_{\mu}B_{R}^{\bar f}
+i\bar {\tilde B}_{L\,\bar f}\gamma^{\mu}\partial_{\mu}\tilde B_{L}^{\bar f}
+i\bar {\tilde B}_{R\, f}\gamma^{\mu}\partial_{\mu}\tilde B_{R}^{\bar f}\\[1mm]
&-m_{B} \Big[\bar {\tilde B}_{R\, f}B_{L}^{f}+\bar B_{R\,\bar f}\tilde B_{L}^{\bar f} +{\rm h.c.} \Big]
\end{split}
\eeq
It shows that the 4-component Dirac spinors are formed by $\{B_{L}^{f},\tilde B_{R}^{f}\}$ and 
$\{\tilde B_{L}^{\bar f}, B_{R}^{\bar f}\}$. The parity conservation is reflected in symmetry
under permutations: $B_{L}^{f}\leftrightarrow B_{R}^{\bar f}$, $\tilde B_{L}^{f}\leftrightarrow \tilde 
B_{R}^{\bar f}$. Thus, $1/2^{+}$ and $1/2^{-}$ 4-component spinors are 
\beq 
B_{+}=\frac{1}{\sqrt{2}}\left(\begin{array}{c}B_{L}+\tilde B_{L} \\ \tilde B_{R}+B_{R} \end{array}\right),
\qquad
B_{-}=\frac{1}{\sqrt{2}}\left(\begin{array}{c}\tilde B_{L}-B_{L}\\ B_{R}-\tilde B_{R} \end{array}\right).
\eeq
In terms of $B_{\pm}$
\beq
{\cal L}_{B}=i\bar B_{+}\gamma^{\mu}\partial_{\mu}B_{+}
+i\bar B_{-}\gamma^{\mu}\partial_{\mu}B_{-}
-m_{B} \bar B_{+} B_{+}-m_{B}\bar B_{-}B_{-}\,,
\eeq
where summation over flavors is implied.

Couplings of spin 0 mesons to baryons  linear in the meson matrix $ M^{f}_{\bar f}$ are
 \beq
 \label{mbb}
 M^{f}_{\bar f}\left[h\,\bar B_{L\,f} B_{R}^{\bar f}+
 \tilde h\,\bar { \tilde B}_{R\,f}  \tilde B_{L}^{\bar f}\right] +{\rm h.c.}\,,
 \eeq
 where two constants $h$ and $\tilde h$ are real to maintain $P$-invariance.
It implies that couplings, e.g.,  to pions written in terms of $B_{\pm}$ are
\beq
\label{piBB}
-i\,\frac{h+\tilde h}{2}\,\big(\bar B_{+}\bm{\tau \pi}B_{-}-\bar B_{-}\bm{\tau \pi}B_{+}\big)
-i\,\frac{h-\tilde h}{2}\,\big(\bar B_{+}\bm {\tau \pi}\,\gamma_{5}B_{+}
-\bar B_{-}\bm {\tau \pi}\,\gamma_{5}B_{-}\big)\,.
\eeq
Cross-couplings linear in both $B$ and $\tilde B$ contain two meson fields, e.g.,
\beq
\label{quad}
M^{f}_{\bar f}\bar M_{g}^{\bar f}\,\bar B_{L\,f} \tilde B_{R}^{g}\,,
\qquad
M^{f}_{\bar f}\partial_{\mu }\bar M_{g}^{\bar f}\,\bar B_{R\,f}\,\gamma^{\mu} \tilde B_{R}^{g}\,.
\eeq 

For spin 1 mesons described by  $V^{L\,f}_{\mu\, g}$, $V^{R\,\bar f}_{\mu\, g}$ couplings are
\beq
V^{L\,f}_{\mu\, g}\left[g\,\bar B_{L\,f}\,\gamma^{\mu} B_{L}^{g}+
 \tilde g\,\bar { \tilde B}_{R\,f}\,\gamma^{\mu} \tilde B_{R}^{g}\right] +
V^{R\,\bar f}_{\mu\,\bar g}\left[g\,\bar B_{R\,\bar f}\,\gamma^{\mu} B_{R}^{\bar g}+
 \tilde g\,\bar { \tilde B}_{L\,\bar f}\,\gamma^{\mu} \tilde B_{L}^{\bar g}\right] +
 {\rm h.c.}
\eeq
In particular for $\rho$ meson it gives
\beq
\frac{g+\tilde g}{2}\,\big(\bar B_{+}\bm{\tau \rho}_{\mu}\,\gamma^{\mu}B_{+}+
\bar B_{-}\bm{\tau \rho}_{\mu}\,\gamma^{\mu}B_{-}\big)
-\frac{g-\tilde g}{2}\,\big(\bar B_{+}\bm {\tau \rho}_{\mu}\,\gamma^{\mu}\gamma_{5}B_{+}
+\bar B_{-}\bm {\tau \rho}_{\mu}\,\gamma^{\mu}\gamma_{5}B_{-}\big)\,.
\eeq
Again cross-coupling containing $B$ and $\tilde B$ are quadratic in meson matrices.

\section*{Appendix B: Spontaneous symmetry breaking \\ and generalized 
Goldberger--Treiman relation}

\renewcommand{\theequation}{B.\arabic{equation}}

\setcounter{equation}{0}

\renewcommand{\thesubsection}{B.\arabic{subsection}}
\setcounter{subsection}{0}

\label{spont}

Spontaneous symmetry breaking can be introduced as nonvanishing vacuum average 
of the meson matrix $M^{f}_{\bar f}$. As we discussed above in the $N_{f}=2$ case this matrix 
contains 8 real fields. There are two U(2)$_{L}\times$U(2)$_{R}$ invariants for this matrix,
${\rm Tr}\, M M^{\dagger}$ and $|{\rm Det}\, M|^{2}$. Correspondingly a generic matrix $M$
can be presented in the form
\beq
M=\sigma\, {\rm e}^{i\eta}\,U\,V\,,
\eeq
where $\sigma$ and $\eta$ are real numbers, the matrix $U\in$ SU(2), i.e. unitary and unimodular, 
like $U=\exp(-i\bm{\tau \pi }/\sigma_{0})$, and 
the matrix $V$ is Hermitian and unimodular, i.e. $V=\exp (\bm{\tau \sigma}/\sigma_{0})$.

If  $\langle V \rangle_{0}\neq \mathbb I$ (i.e. $\langle \sigma_{3} \rangle_{0}\neq 0$) then
 U(2)$_{L}\times$U(2)$_{R}$ symmetry is spontaneously broken to 
U(1)$_{B}\times$U(1)$_{I_{3}}$ and six Goldstone bosons appear:  triplet of pseudoscalars,
$\bm \pi$, singlet pseudoscalar, $\eta$, and two scalars, $\sigma^{\pm}=(\sigma_{1}\mp
i\sigma_{2})/\sqrt{2}$. Such a pattern of spontaneous breaking 
when the vector SU(2)$_{V}$ is broken  is not allowed in QCD \cite{Witten}, it means that  we should
take  $\langle V \rangle_{0}= \mathbb  I$.  

Then the pattern of breaking is standard,
U(2)$_{L}\times$U(2)$_{R}\to$U(1)$_{B}\times$SU(2)$_{V}$, with four Goldstones, $\bm \pi$ and 
$\eta$. The fourth one, $\eta$, is actually pseudo-Goldstone because the associated U(1)$_{A}$ symmetry is anomalous in QCD. Still we have to keep $\eta$, together with scalars $\sigma$ and 
$\bm \sigma$, as massive partners of pions when considering asymptotically linear realization
of the chiral symmetry.

To see how the spontaneous breaking acts on baryons let us substitute $M$ in Eq.\,\eqref{mbb}
by its vacuum average $\langle M \rangle_{0}=\sigma_{0}\mathbb I$. This lifts degeneracy 
in masses of $1/2^{+}$ and $1/2^{-}$ baryons,
\beq
m_{B_{-}}-m_{B_{+}}=\sigma_{0}(h+\tilde h)\,.
\eeq
Equation \eqref{piBB} shows that the same combination $h+\tilde h$ enters
the $\pi B_{-}B_{+}$ coupling,
\beq
g_{\pi B_{-}B_{+}}=-i\,\frac{h+\tilde h}{2}\,,
\eeq
so 
\beq
\label{gotr1}
m_{B_{-}}-m_{B_{+}}=2i\sigma_{0}\, g_{\pi B_{-}B_{+}}\,.
\eeq
This equation is an analogue of the Goldberger--Treiman relation.

 \begin{figure}[h]
 \centerline{\includegraphics[width=2in]{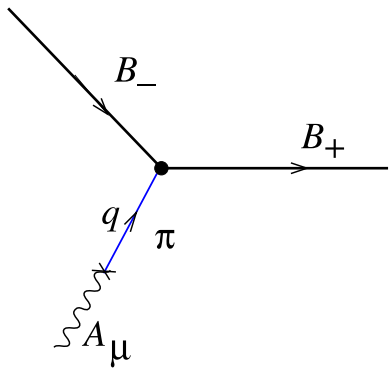}}
 \caption{\small The pion pole in the  $B_-$ to $B_+$ matrix element of the axial current}
 \label{pd}
 \end{figure}

Indeed, let us sandwich the axial current $A^{3}_{\mu}$ between  upper isospin components of
 $B_{\pm}$,
\beq
\langle B_+ | 2A^{3}_\mu |B_-\rangle =
\Big(
g_{\mu\nu} - \frac{q_\mu q_\nu}{q^2}
\Big)\bar B_+\gamma^\nu B_-\,,
\label{msbj}
\eeq
see Fig.\,\ref{pd}  for notation. The expression in parentheses makes the axial current matrix
element explicitly transverse, as is required by the axial current conservation in the
chiral limit. Here we dropped an overall constant 
in front of $\bar B_+\gamma^\mu B_-$ (an analogue of $g_A$)
as it is expected to be 1 in the limit of heavy $B_{\pm}$.\footnote{%
\,Deviations from 1 are reflected in additions to $\pi B_{-}B_{+}$ vertex
due to the couplings \eqref{quad}. The second of these couplings
containing derivatives contributes after symmetry breaking.} 

Comparing the residue of the pole from Eq.\ (\ref{msbj}) and Fig.\,\ref{pd}
we immediately conclude that
\beq
2ig_{\pi B_{-}B_{+}}\, F_\pi = m_{B_{-}}-m_{B_{+}}\,.
\label{msbj1}
\eeq
This is the same relation as Eq.\ \eqref{gotr1} with the identification
$\sigma_{0}=F_{\pi}$.
 Equation (\ref{msbj1}) implies that for the 
$n$-th ``radial" excitation $(\Delta m_\pm)_n$ scales as $(g_{\pi B_{-}B_{+}})_n$,
and falls off with $n$ with the same rate. 

We can apply a similar consideration to coupling of the axial-vector boson $a_{1}$ to  $B_{-}B_{+}$,
its contribution to the nonpole part of the matrix element \eqref{msbj} is given by the same Fig.\,\ref{pd}
with the substitution of $\pi$ by $a_{1}$.  In this way we arrive at
\beq
\label{aBB}
\frac{2F_{a_{1}} \,g_{a_{1}B_{-}B_{+}}}{m^{2}_{a_{1}}}=-1\,,
\eeq
where $F_{a_{1}}$ is the coupling of $a_{1}$ to the axial current and $m_{a_{1}}$ is its mass.
The relation \eqref{aBB} demonstrates clearly that $g_{a_{1}B_{-}B_{+}}$ does not decrease 
for high excitations in contrast with $g_{\pi\,B_{-}B_{+}}$.

\end{document}